\documentstyle[twocolumn,aps]{revtex}

\begin{document}

\draft
\title{ Critical behavior in 2+1 dimensional black holes}
\author{ Rong-Gen Cai\footnote{E-mail: cairg@itp.ac.cn},
 Zhi-Jiang Lu and Yuan-Zhong Zhang }
\address{CCAST (World Laboratory), P.O. Box 8730, 
Beijing 100080, China \\
and Institute of Theoretical Physics, Academia 
Sinica, P.O. Box 2735, 
Beijing 100080, China}
\maketitle

\begin{abstract}
The critical behavior and phase transition in 
the 2+1 dimensional Ba\~nados, Teitelboim, and Zanelli (BTZ) black
 holes are discussed. By calculating the equilibrium thermodynamic
 fluctuations in the microcanonical ensemble, canonical ensemble, and 
grand canonical ensemble, respectively, we find that the extremal 
spinning BTZ black hole is a critical point, some critical exponents
 satisfy the scaling laws of  the ``first kind'', and the scaling laws
 related to the correlation length  suggest  that the effective spatial
 dimension of extremal black holes is one, which is in 
  agreement with the argument that the extremal black holes  are the
 Bogomol'nyi saturated string states. In addition, we find that the 
massless BTZ black hole is  a critical point of spinless BTZ black
 holes. 
\end{abstract}

\section{Introduction}

In 1973, Bardeen, Carter, and Hawking [1] established a remarkable 
mathematical analogy existed between the laws of thermodynamics and the
 laws of black hole 
mechanics derived from general relativity. If the formal replacements 
$E\rightarrow M$, $T\rightarrow c\kappa$, and $S\rightarrow A/8\pi c$ 
( $c$ is a constant) are made in the laws of thermodynamics, one can obtain 
immediately the laws governing the mechanics of black holes, where $E$, 
$T$, and $S$ are the internal energy, temperature, and entropy of an 
ordinary thermodynamic system, 
while the $M$, $\kappa $, and $A$ denote the mass, surface gravity, and 
horizon area of a black hole, respectively. The physical analogy, 
however, seems to have problems due to the fact that in classical 
general relativity the  thermodynamic temperature of a black hole 
should be absolute zero. When  quantum effects are taken into account,
 Hawking [2] found that a black hole absorbs and emits particles as a 
black body at the temperature $T=\kappa/2\pi$. The discovery of Hawking
 resolved the puzzle of this physical analogy and 
    provides a basis of the  black hole entropy suggested firstly by
 Bekenstein [3]. Since then, most people have believed that a black hole is
 a thermodynamic system, and the thermodynamics of black holes has been 
developed [4,5].

 The phase transition is an important phenomenon in the ordinary 
thermodynamics.  Therefore, it is natural to ask whether the phase 
transitions exist in the black  hole thermodynamics. According to the 
infinite discontinuity of the heat capacity of Kerr-Newman black holes, 
Davies [4] claimed that a second-order phase
 transition  takes place when the black holes leap over the discontinuity
 point of heat capacity.  Although some researches on this singularity 
point have been carried out [6,7,8,9,10],
   and Lousto [9] even claimed that the phase transition of Davies satisfies
 the scaling laws of critical points, the reasons that the Davies' point is
 regarded as a phase transition point are  insufficient, because the event 
horizon does not lose its regularity and the internal state of the black
 hole is unaffected at those 
points.  Furthermore, by investigating the stability of black holes in the 
different  environments, Kaburaki and his collaborators [11,12] found that 
the Davies' points are in fact related to the presence of the changes of 
stability, and further claimed that the divergence of heat capacity at
 those points does not mean the occurrence of phase transitions.

 On the other hand, more than a decade ago, based on the analogy between 
the Kerr  black hole and a laser system, Curir [13] claimed that a critical
 point exists 
 at the extremal limit of the hole and a phase transition takes place from
 the  extremal to nonextremal Kerr black hole. By making use of
 Landau-Lifshitz theory 
 of nonequilibrium thermodynamic fluctuations, Pav\'on and Rub\'\i\  [14] 
calculated,
  respectively, some second moments of relevant quantities for Schwarzchild
   black
   holes, Kerr black holes, and Reissner-Nordstr\"{o}m (RN) black holes, 
   and found 
   that some second moments diverge for the extremal Kerr black holes and 
   extremal  RN black holes, but are always finite for Schwarzchild black 
   holes. The most 
   important thing is that nothing special happens at the Davies' points. 
   The 
   divergence of second moments leads to the argument of Pav\'on and 
   Rub\'\i\  that a
    second-order phase transition occurs from the extremal to nonextremal 
    black holes.
 This conclusion is in agreement with the one of Curir although the two 
 methods they employed  are different.

To find the essence and universality of divergence of second moments, 
Cai, Su and Yu 
discussed the nonequilibrium and equilibrium fluctuations for various 
charged dilaton black holes [15], 
and found the difference of the outer and inner horizons plays a crucial 
role in the 
divergence of second moments. Thus, they suggested that the difference plays
 the role of an
 order parameter in the second-order phase transition of black holes [16]. 
 More recently, 
 Kaburaki [17] has shown that the critical exponents in the extremal
  Kerr-Newman black 
 holes also obeys the scaling laws. As is well known, the extremal black
  hole is very
  different from the nonextremal one, for example, in the geometric 
  structure, 
  thermodynamics [18,19], the essence of radiations, etc.. 
  Wilczek {\it et al} [20]
   argued that the thermodynamic description is inadequate for 
   some extremal black 
   holes and the behavior of these holes resembles the normal elementary
    particles' or strings' behavior.

To investigate the properties of black holes, one of the powerful methods 
is to
 study the interaction between quantum fields and black holes. Traschen [21] 
has 
 recently finished such a issue. By studying the behavior of a massive 
 charged scalar
  field on the background of RN black holes, Traschen has found that the 
  spacetime
   geometry near the horizon of the extremal RN black holes has a scaling 
   symmetry,
    which is absent for the nonextremal holes, a scale being introduced by
     the
     surface gravity. The scaling symmetry results in that an external
      source has a
      long range influence on the extremal background, compared to a 
      correlation 
      length scale which falls off exponentially fast in the case of 
      nonextremal holes.
       The long range correlation is just the characteristic of phase 
       transitions. 
       Although these evidences are in favor of the existence of phase 
       transitions in the extremal black holes, further study is needed
        in order to better understand the behavior of the critical point
         because the Hawking
          temperature of an extremal black hole vanishes and the ordinary 
          thermodynamics is invalid in this limit.

In recent years, one of the important progresses in general relativity is 
the
 discovery of the 2+1 dimensional black holes by Ba\~nados, Teitelboim, 
 and Zanelli
  (BTZ) [22]. A review paper on the 2+1 dimensional black holes can be seen
   in
   Ref. [23]. Although the BTZ black holes  are the counterparts of Kerr
    black 
   holes in 2+1 dimensions, their thermodynamic properties have some 
   differences.
    An important difference is that the heat capacity of BTZ black holes
     is always 
    positive and has no discontinuous jump, while the heat capacity of Kerr
     holes 
    is negative for the small angular momentum, and positive for the large 
    angular 
    momentum, that is, it has an infinite discontinuity at the Davies' point.
     This
     property results in that not only the microcanonical ensemble, but also
      the 
     canonical ensemble can describe the gas of BTZ black holes, and no 
     limited 
 temperature exists [24], unlike the gas of the 1+1 dimensional dilaton
  black holes, which has a limited temperature [25].

  The purpose of this paper is to investigate the equilibrium thermodynamic 
  fluctuations of BTZ black holes in the different enviroments and the
   behavior
   of critical points. We find that similar to the Kerr black holes, a 
      critical
 point exists in the extremal BTZ black holes. Furthermore, the massless 
 BTZ black hole is found to be a critical point as well, which is different
  from the case of Schwarzchild  black holes. In particular, we find that
   the effective spatial dimension of extremal black holes is one, which 
   is  surprisingly in agreement with the arguments that the extremal
    black holes are the Bogomol'nyi saturated string states and the 
    extremal black holes can be regarded as elementary particles like 
    states in  string theory [26].

The paper is organized as follows. In the next section we will briefly
 review the BTZ black holes and introduce the fluctuation theory of 
 equilibrium thermodynamics. In Sec. III we will discuss the 
 equilibrium fluctuations of thermodynamic quantities for the 
 spinning BTZ black holes in the microcanonical ensemble, canonical
  ensemble, and grand canonical  ensemble, respectively, study the 
  critical behavior of extremal BTZ black holes, and calculate some
   critical exponents of relevant quantities. We will discuss the 
   case of the spinless BTZ black holes, in Sec. IV, and argue that
    the massless BTZ black hole is a critical point of the spinless
     BTZ black holes, and a second-order phase transition takes 
     place from the massless BTZ black hole to a generic spinless
      BTZ black hole. The conclusion and discussion are included 
      in Sec. V.

\section{BTZ black holes and the theory of equilibrium fluctuations}

In 2+1 dimensions, the Einstein gravity theory with a negative
 cosmological constant has the BTZ black hole solution, which 
 can be described by the line element [22]
\begin{equation}
ds^2=-N^2dt^2+N^{-2}dr^2+r^2(N^{\phi}dt +d\phi )^2
\end{equation}
where
\begin{equation}
N^2(r)=-M+\frac{r^2}{l^2}+\frac{J^2}{4r^2},
\ \  N^{\phi}(r)=-\frac{J}{2r^2},
\end{equation}
$M$ and $J$ are two integration constants and represent the mass and
 angular momentum of the BTZ black holes (here the unit $8G=1$ is used),
  respectively, and $l^{-2}$ denotes the cosmological constant. 

The horizons are given by the condition that the lapse function $N^2(r)=0$,
 and read
\begin{equation}
r_{\pm}=\left [\frac{M}{2}l^2(1\pm \triangle)\right]^{1/2}, 
\ \ \triangle= [1-(J/Ml)^2]^{1/2}.
\end{equation}
In order that solution (1) has the structure of black  holes, 
evidently, we must 
impose the condition
\begin{equation}
M\ge 0 \ \ {\rm and}\ \ J\le Ml.
\end{equation}
The BTZ black hole solution (1) is neither asymptotically flat nor 
asymptotically de Sitter, but asymptotically anti-de Sitter.
 If $M=-1$ and $J=0$, the 2+1 dimensional anti-de Sitter space is
  restored
\begin{equation}
ds^2_{\rm ADS}=-(1+\frac{r^2}{l^2})dt^2+(1+\frac{r^2}{l^2})^{-1}dr^2+
r^2d\phi ^2.
\end{equation}
The BTZ black hole solution can be obtained by identifying points in 
this anti-de Sitter space (5) and using the orbits of a spacelike
 Killing vector field. In the BTZ black hole solution (1), a special
  case exists. That is, if $M=0$ and $J=0$, the solution (1) then 
  reduces to 
\begin{equation}
ds^2_{\rm VAC}=-\frac{r^2}{l^2}dt^2+\frac{l^2}{r^2}dr^2 +r^2d\phi ^2,
\end{equation}
which is the ground state of the BTZ solutions, and describes a 
massless black hole. The massless BTZ solution (6) is of some 
interesting properties. First, the solution (6) has an infinitely 
long throat for small $r>0$, which is reminiscent of the extremal
 RN black hole solution in 3+1 dimensions. This leads the solution 
 (6) to have similar properties as those of the extremal RN black 
 holes. Second, the solution (6) has two exact supersymmetries, 
 while a generic extremal BTZ black hole has only one supersymmetry [27].

Define
\begin{equation}
\Omega _H\equiv\left. -\frac{g_{t\phi}}{g_{\phi\phi}}\right |_{r_+}
=\frac{J}{2r_+^2},
\end{equation}
analogous to its counterpart in the Kerr black holes, $\Omega _H$ 
can be regarded as the angular velocity of the BTZ black holes. In 
addition, the surface $r^s$ is the surface of infinite redshift 
where $g_{tt}(r^s) $ vanishes,
\begin{equation}
r^s=\sqrt{M}l.
\end{equation}
Obviously, $r^s\ge r_+$. The region
\begin{equation}
r_+\le r\le r^s
\end{equation} 
is called as the ergosphere of the BTZ black hole. Because of the 
appearance of the ergosphere, the Penrose process and Misner 
superradiation can take place in the BTZ black holes, even under 
the extremal condition ($J=Ml$). When $J=0$, the ergosphere 
disappears and the superradiation thereby is killed.

By analyzing the geometry of solution (1), one can find that it has 
two Killing vectors, timelike  $k^{\mu}_{(t)}$ and spacelike
 $m^{\mu}_{(\phi )}$. Hence,
\begin{equation}
l^{\mu}=k^{\mu}_{(t)}+\Omega _H m^{\mu}_{(\phi)}
\end{equation}
is also a Killing vector. According to the definition of surface
 gravity, $l^{\mu}_{;\nu}l^{\nu}=\kappa l^{\mu}$, we obtain the 
 surface gravity of BTZ black holes at the horizon $r_+$,
\begin{equation}
\kappa =\frac{r_+^2-r_-^2}{r_+l^2}=\frac{M\triangle}{r_+}.
\end{equation}
Furthermore, making use of the definition of mass,
\begin{equation}
M=\frac{1}{4\pi}\int _{r_+}k^{\mu;\nu}_{(t)}d\Sigma _{\mu\nu},
\end{equation}
we can get the Smarr-type mass formula of BTZ black holes,
\begin{equation}
M=\frac{\kappa}{2\pi}L+\Omega _HJ,
\end{equation}
where $L=2\pi r_+$ is the perimeter of the horizon $r_+$. The Hawking 
temperature of the black holes is
\begin{equation}
T=\frac{\kappa}{2\pi}=\frac{r_+^2-r_-^2}{2\pi r_+l^2}.
\end{equation}
With the aid of Euclidean action method, one can easily obtain the 
black hole entropy [22]
\begin{equation}
S=4\pi r_+=4\pi \left[\frac{M}{2}l^2(1+\triangle)\right ]^{1/2},
\end{equation}
which is the twice perimeter of the horizon. Varying the mass formula
 (13) yields the first law of thermodynamics for the BTZ black holes
\begin{equation}
dM=TdS+\Omega _H dJ.
\end{equation}
In terms of the formula $C_J=\left (\frac{\partial M}{\partial T}
\right )_J$, we get the heat capacity of the hole,
\begin{equation}
C_J=\frac{4\pi r_+\triangle}{2-\triangle}.
\end{equation}
Because of $0\le \triangle \le 1$, $C_J\ge 0$ is always satisfied. 
It is very different from that of Kerr black  holes [4]. When $J=0$,
 i.e., $\triangle =1$, one has $C_J\equiv C=4\pi l\sqrt{M}$, which 
 means the temperature to increase with the mass , and to $T=C=0$ as
  $M=0$. When $J=Ml$, i.e., $\triangle =0$, one has $C_J=0$, which 
  corresponds to the extremal BTZ black holes, and the vanishing 
  Hawking temperature.

We now turn to the fluctuation theory of equilibrium thermodynamics. 
In general, the fluctuations of thermodynamic quantities of a 
thermodynamic system equilibrating with different environments
 are different. In some textbooks on thermodynamics [28-30],
  however, this point is not distinguished clearly. In order 
  that the Einstein's theory of thermodynamic fluctuations is
   applicable in the various thermodynamic environments, 
   Kaburaki [12] developed the Einstein's theory and made 
   it be convenient to discuss the fluctuations in various
    ensembles.

The equilibrium states of a thermodynamic system in a given 
environment are characterized by the Massieu function, $\Phi$,
 and the infinitesimal variation of the function $\Phi$ can 
 be written as [12]
\begin{equation}
d\Phi=\sum^{n}_{i=1}X_idx_i,
\end{equation}
where the variables $\{x_i\}$ are called the intrinsic variables 
and are specified directly by the given thermodynamic environment. 
The $\{X_i\}$ forms a set of variables conjugate to $\{x_i\}$, and 
can be expressed as functions of their intrinsic variables:
 $X_i=(\partial \Phi/\partial x_i)_{\bar{x}_i}$, where $\bar{x}_i$ 
 denotes the set of variables $\{x_i\}$ eliminating the $x_i$. Thus
  the probability of the system deviating the equilibrium state can
   be given by the macroscopic distribution function
\begin{equation}
P(\xi _1\ldots \xi _n)d\xi _1\ldots d\xi _n\sim \exp [k_B^{-1}
(\hat{\Phi}-\Phi)]d\xi _1\ldots d\xi _n,
\end{equation}
where $k_B$ is the Boltzmann's constant, $\xi _i\equiv \delta X_i$ 
stands for the deviation of the $i$th conjugate variable from its 
equilibrium value and $\hat{\Phi}$  represents the Massieu function 
analytically continued to the nonequilibrium points near the 
equilibrium sequence in the phase space. By expanding the Massieu
 function with respect to conjugate variables up to the second order,
  one has
\begin{equation}
P(\{\xi _i\})\sim \exp[-(2k_B)^{-1}\sum^{n}_{i=1}\lambda _i\xi ^2_i],
\end{equation}
where $\lambda _i=(\partial x_i/\partial X_i)_{\bar{x}_i}=(\partial ^2 
\Phi/\partial x^2_i)^{-1}_{\bar{x}_i}$ are the eigenvalues of the 
fluctuation modes $\delta X_i$. From (20), as is expected the average 
of each fluctuation mode always vanishes, $<\delta X_i>=0$, and the 
second moments of equilibrium fluctuations are
\begin{equation}
<\delta X_i\delta X_j>=\frac{k_B}{\lambda _i}\delta _{ij}=k_B\left 
(\frac{\partial ^2 \Phi}{\partial x^2_i}\right)_{\bar{x}_i}\delta _{ij}.
\end{equation}
From (21) we can see that the second moments diverge only if 
$\lambda _i=0$, i.e., $(\partial X_i/\partial x_i)_{\bar{x}_i}
=\pm \infty$. According to the turning point method of stability
 analysis developed by Katz [31], the change of stability occurs 
 only at the turning point where the tangent of the curve changes
  its sign through a infinity (a vertical tangent) in the 
  ``conjugate diagrams'', in which the conjugate variables 
  $\{X_i\}$ are plotted against a intrinsic variable (other
   instrinsic variables are fixed). The Davies' points are 
   just such turning points. The divergence of some second 
   moments at the Davies' points only means the change of 
   stability for Kerr-Newman black holes, and does not have 
   close relation to the second-order phase transitions [11,12],
    Instead the divergence of some second moments at the extremal
     limit might imply the presence of phase transitions because 
     the extremal limit is not the turning point.

\section{Fluctuations of spinning BTZ black holes and scaling laws}

In this section, by using the fluctuation theory developed in the 
previous section, we will discuss the equilibrium fluctuations of 
spinning BTZ black holes in the microcanonical ensemble, canonical 
ensemble, and grand canonical ensemble, respectively, and find that 
all the second moments are finite in these ensembles except for the 
extremal BTZ black holes. The extremal limit is a critical point 
and critical exponents obey the scaling laws. The effective spatial
 dimension of extremal black holes is found to be one. The spinless
  BTZ black holes will be discussed separately in the next section.

\subsection{Equilibrium fluctuations of spinning BTZ black holes}

The proper Massieu functions are different for different environments. 
Therefore, 
the corresponding fluctuations and second moments are different in the
 various ensembles. Let us discuss the case of microcanonical 
 ensemble first.

(1) Microcanonical ensemble. In this ensemble, nothing of the 
thermodynamic system can be exchanged with its environment. The 
proper Massieu function is the entropy of system. For the BTZ 
black holes, rewriting the first law of BTZ black holes (16), 
we have
\begin{equation}
d\Phi _1=dS=\beta dM-\mu dJ,
\end{equation}
where $\beta=1/T$ and $\mu=\beta \Omega _H$. Thus, in this ensemble, 
the intrinsic variable are $\{M,J\}$ and the corresponding conjugate 
variables are $\{\beta, -\mu\}$. The eigenvalues for fluctuation modes
 are
\begin{eqnarray}
\lambda _{ 1m}&=&\left (\frac{\partial M}{\partial \beta} \right )_J
=-T^2C_J,\\
\lambda _{ 1j}&=&-\left (\frac{\partial J}{\partial \mu}\right )_M=
-TI_M,
\end{eqnarray}
where $C_J$ is the heat capacity given by Eq. (17), and 
\begin{eqnarray}
I_M & \equiv & \beta \left (\frac{\partial J}{\partial \mu}\right )_M 
,\nonumber\\
&=&\left [\frac{1}{2r_+^2}+\frac{J^2}{8Mr_+^4\triangle}
+\frac{J^2}{2M^2l^2r_+^2\triangle}\right ]^{-1},
\end{eqnarray}
is the ``moment of inertia'' of BTZ black holes. Some second moments
 can be easily given,
\begin{eqnarray}
<\delta \beta \delta \beta>&=&-k_B\frac{\beta ^2}{C_J},\nonumber\\
<\delta \mu \delta \mu>&=&-k_B\frac{\beta}{I_M},\nonumber\\
<\delta \Omega _H\delta \Omega _H>&=&-k_B\left [\frac{T}{I_M}
+\frac{\Omega ^2_H}{C_J}\right ], \nonumber\\
<\delta \Omega _H\delta \beta >&=&k_B\frac{\beta \Omega _H}{C_J}.
\end{eqnarray}
Obviously, these second moments are finite for the nonextremal
 BTZ black holes. When the extremal limit is approached, i.e., 
 $\triangle \rightarrow 0$, however, the eigenvalues $\lambda _{1m}$ 
 and $\lambda _{1j}$ approach zero, and all of these second moments 
 (26) diverge. The divergence means that the extremal BTZ black hole
  is a critical point.

(2) Canonical ensemble. In this ensemble, the BTZ black hole can only
 exchange the heat with surroundings. The variation of the proper 
 Massieu function is
\begin{eqnarray}
d\Phi _2&=&dS-d(\beta M),\nonumber\\
        &=&-Md\beta -\mu dJ,
\end{eqnarray}
which gives the intrinsic variables $x_i=\{\beta, J\}$ and the 
conjugate variables $X_i=\{-M,-\mu\}$. The corresponding
 eigenvalues are 
\begin{eqnarray}
\lambda _{2\beta}&=&-\left (\frac{\partial \beta }
{\partial M}\right )_J =\frac{\beta ^2}{C_J},\\
\lambda _{2j}&=&-\left (\frac{\partial J}{\partial \mu}
\right )_{\beta}=-\frac{I_{\beta}}{\beta},
\end{eqnarray}
where
\begin{eqnarray}
I_{\beta}&\equiv &\beta \left (\frac{\partial J}{\partial 
\mu}\right)_{\beta},\nonumber\\
&=&\left[\frac{1}{2r_+^2}+\frac{J}{r_+^2}\left (\frac{J}{M^2l^2
\triangle}\right.\right.\nonumber\\
&-&\left.\left.\left (\frac{1}{M}+\frac{J^2}{M^3l^2\triangle ^2}
\right ){\cal G}\right )\right ]^{-1},
\end{eqnarray}
and
\begin{eqnarray}
{\cal G}&=&\frac{J}{Ml^2\triangle}\left(\frac{Ml^2}{4r_+}-
\frac{r_+}{\triangle}\right)\left [\frac{(1+\triangle)l^2}{4r_+}
\right.\nonumber\\
&-&\left.\frac{r_+}{m}+\frac{J^2}{M^3l^2\triangle }\left(
\frac{Ml^2}{4r_+}-\frac{r_+}{\triangle}\right)\right ]^{-1}.
\nonumber
\end{eqnarray}
With the help of Eqs. (20) and (21), in this ensemble, some
 nonvanishing second moments are
\begin{eqnarray}
<\delta M\delta M>&=&k_BT^2C_J,\nonumber\\
<\delta S\delta M>&=&k_BTC_J,\nonumber\\
<\delta \mu\delta\mu>&=&-k_B\frac{\beta}{I_{\beta}},\nonumber\\
<\delta \Omega _H\delta \Omega _H>&=&-k_B\frac{T}{I_{\beta}}, \nonumber\\
<\delta S\delta S>&=&k_B C_J.
\end{eqnarray}
These second moments are also finite for nonextremal BTZ black holes. 
In the extremal limit, $\lambda _{2\beta}$ diverges, but $\lambda _{2j}$
 does not. Thus, $<\delta M\delta M>$, $<\delta S\delta M>$, and $<\delta
  S\delta S>$ are finite, but $<\delta \mu\delta\mu>$ and $<\delta 
  \Omega _H\delta \Omega _H>$ diverge.

(3) Grand canonical ensemble. This ensemble implies that the BTZ black 
holes not only can exchange the heat with surroundings, but also do 
work to the surroundings. The variation of proper Massieu function 
in this case is 
\begin{eqnarray}
d\Phi _3&=&d\Phi _2+d(\mu J)\nonumber\\
        &=&-Md\beta +Jd\mu.
\end{eqnarray}
 The intrinsic variables become $x_i=\{\beta, \mu\}$, and the 
 corresponding conjugate variables $ X_i=\{-M, J\}$. The eigenvalues  
 of fluctuation modes are
\begin{eqnarray}
\lambda _{3\beta}&=&-\left(\frac{\partial \beta}{\partial M}\right
 )_{\mu}=\frac{\beta ^2}{C_{\mu}},\\
\lambda _{3\mu}&=&\left(\frac{\partial \mu}{\partial J}\right)_{\beta}
=\frac{\beta}{I_{\beta}},
\end{eqnarray}
where 
\begin{eqnarray}
C_{\mu}&\equiv &\left(\frac{\partial M}{\partial T}\right)_{\mu},
\nonumber\\
&=&2\pi\left\{\frac{\triangle}{2r_+}+\left (\frac{J}{Mr_+\triangle}
+\frac{JMl^2}{4r_+^3\triangle}\right)^{-1}\right.\nonumber\\
& &\left (\frac{M}{r_+}-\frac{M^2l^2\triangle}{4r_+^3}\right)\left [
\frac{1}{Mr_+\triangle}{\cal H}\right.\nonumber\\
&-&\left. \left. \left(\frac{J}{M^2r_+\triangle}+\frac{l^2(1+
\triangle)J}{4Mr_+^3\triangle}
\right) \right] \right\}^{-1},
\end{eqnarray}
and
\begin{eqnarray}
{\cal H}&=&\left [\frac{J}{M^2r_+}+\frac{l^2(1+\triangle)J}{4Mr_+^3}
\right.\nonumber\\
&+&\left.\frac{J^2}{M^2l^2}\left(\frac{J}{Mr_+\triangle ^2}+\frac{JMl^2}
{4r_+^3\triangle}\right) \right]\nonumber\\
& & \left [\frac{1}{Mr_+}+\frac{J}{M^2l^2}\left (\frac{J}{Mr_+
\triangle ^2}+\frac{JMl^2}{4r_+^3\triangle}\right )\right ]^{-1}.
\nonumber
\end{eqnarray}
The nonvanishing second moments are easily given and read
\begin{eqnarray}
<\delta M\delta M>&=&k_BT^2 C_{\mu},\nonumber\\
<\delta J\delta J>&=&k_BT I_{\beta}, \nonumber\\
<\delta S\delta S>&=&k_B(C_{\mu}+\mu ^2 T I_{\beta}),\nonumber\\
<\delta S\delta M>&=&k_BT C_{\mu},\nonumber\\
<\delta J\delta S>&=&-k_B\mu ^2T I_{\beta}.
\end{eqnarray}
In this case, the eigenvalues $\lambda _{3\beta}$ and $\lambda _{3\mu}$
 diverge under the extremal condition, but, all of these second moments 
 are finite and vanish for the extremal BTZ black holes.

\subsection{Scaling laws at the critical point}

 To discuss the critical behavior of an isolated black holes, it is 
 appropriate to choose the microcanonical ensemble, in which the 
 proper Massieu function is the entropy of black holes. From the 
 previous subsection we see that in the microcanonical ensemble, 
 the eigenvalues of fluctuation modes are zero and nonvanishing 
 second moments diverge for extremal BTZ black holes. Therefore, 
 the extremal limit is a critical point of BTZ black holes.  The 
 extremal and nonextremal BTZ black holes are two different phases. 
 In the ordinary thermodynamic system, the order parameters are
  defined usually as the differences of some extensive quantities 
  between two phases. Due to the differences between black holes 
  and ordinary thermodynamic system, however, the corresponding
   order parameters in black hole thermodynamics are defined as
    the differences of conjugate variables between the two phases.
     This is because that the intensive quantities are different 
     in the two phases while extensive ones are common. 
      Correspondingly, Kaburaki [17] defined various critical 
      exponents associated with some thermodynamic quantities 
      of Kerr-Newman black holes. For the BTZ black holes, in 
      this case, the instrinsic variables $x_i=\{ M, J\}$, and
       the conjugate variables $X_i=\{\beta, -\mu \}$. Thus,  
       $\eta _{\beta}=\beta _+-\beta _-$ and $\eta _J=\mu_+-\mu _-$ 
       can be regarded as the order parameters of BTZ black holes, 
       where the suffixes ``$+$'' and ``$-$'' mean that the quantity 
       is taken at the outer horizon and inner horizon, respectively.
        The second-order derivatives of entropy with respect to the
         intrinsic variables are the inverse eigenvalues,
\begin{eqnarray}
\bar{\zeta}_1&\equiv &\left (\frac{\partial ^2S}{\partial M^2}\right)_J
=\lambda _{1m}^{-1}=-\frac{\beta ^2}{C_J},\\
\bar{\zeta}_2&\equiv&\left (\frac{\partial ^2 S}{\partial J^2}\right)_M
=\lambda _{1j}^{-1}=-\frac{\beta}{I_M}.
\end{eqnarray}
According to the definition of Kaburaki [17], some critical exponents 
can be given as
\begin{eqnarray}
\bar{\zeta}_1&\sim&\varepsilon _M^{-\alpha}\ \ \ ({\rm for\ J\ fixed}),
\nonumber\\
         &\sim&\varepsilon _J^{-\varphi}\ \ \ ({\rm for\ M\ fixed}),\\
\bar{\zeta}_2 & \sim & \varepsilon ^{-\gamma}_M \ \ \  ({\rm for\ J\ fixed}),
 \nonumber\\
          & \sim &\varepsilon ^{-\sigma}_J \ \ \ ({\rm for\ M\ fixed}),\\
\eta _J&\sim &\varepsilon ^{\beta}_M \ \ \ ({\rm for\ J\ fixed}),\nonumber\\
      &\sim &\varepsilon ^{\delta ^{-1}}_J \ \ \ ({\rm for\ M\ fixed}),
\end{eqnarray}
where $\varepsilon _M$ and $\varepsilon _J$ represent the infinitesimal
 deviations of $M$ and $J$ from their limit values. From Eqs. (39)-(41),
  these critical exponents are easily obtained,
\begin{equation}
\alpha=\varphi=\gamma=\sigma=3/2,\ \ \beta=\delta ^{-1}=-1/2.
\end{equation}
These values are exactly the same as those of Kerr-Newman black holes [17]. 
It shows the universality of critical behavior for extremal black holes. 
Naturally, those critical exponents satisfy the scaling laws of the  
``first kind'',
\begin{eqnarray}
&&\alpha +2\beta +\gamma =2,\\
&&\beta(\delta-1)=\gamma,\\
&&\varphi (\beta +\gamma)=\alpha.
\end{eqnarray}
The scaling laws are related to the fact that the entropy (15) of BTZ
 black holes is a homogeneous function,
\begin{equation}
S(\lambda M, \lambda J)=\lambda ^{1/2}S(M,J),
\end{equation}
where $\lambda$ is a positive constant. Because the extremal black holes 
separate the nonextremal black  holes from the naked singularities and 
the thermodynamic knowledge for the naked singularity is still lacking, 
there exist some shortcomings in these calculations. For example, these 
critical exponents can 
be  computed only in one side, and the ordinary thermodynamics is 
invalid in the extremal limit.   Despite these drawbacks, the 
differences between extremal and nonextremal black  holes seem to
 be enough to show that the extremal limit is a critical point and 
 a second-order phase transition occurs when the extremal black 
  holes become the nonextremal black holes.

To the best of our knowledge, there exist at least three aspects of
 differences between extremal and nonextremal black holes: (1) 
 Symmetry and geometric structure. As is well known, the extremal
  RN black holes are the solutions of the $N=2$ supergravity, and 
  so are of the supersymmetry [32,33]. In fact, many extremal black
   holes are also of the supersymmetry [34]. More recently, Coussaert
    and Henneaux [27] have proved that the supersymmetry also exists 
    in the extremal BTZ black holes. But these supersymmetric properties
     are absent for nonextremal black holes. According to the general
      arguments to second-order phase transitions, we know that the 
      presence of phase transition is always accompanied by the change
       of symmetry of a thermodynamic system. Therefore, The occurrence
        of a phase transition at the extremal black holes is in 
        agreement with the change of symmetry of black holes. 
         The extremal black holes are in the disordered phases, 
         while the nonextremal black holes are in the ordered phases. 
         In addition, it is known that the differences of topological 
         structures exist in the extremal and nonextremal black holes 
         [18,19]. For example, in the Euclidean manifolds of black hole
          solutions, a conical singularity exists at the event horizon 
          for a nonextremal black hole and results in the periodicity of
           the Euclidean time in order to remove the singularity, which 
           gives the Hawking temperature of the black hole. The 
           periodicity is absent for the extremal black holes. This
            gives rise to the argument that an extremal black hole 
            can be in equilibrium with a heat bath at an arbitrary 
            temperature [18]. (2) The difference in thermodynamics.
             A nonextremal black hole can be described very well by
              thermodynamics, but the thermodynamic description is 
              inadequate for some extremal black holes [20]. 
(3) The kinds of radiations. As is well known, the nonextremal black 
holes have both the thermal Hawking evaporation and the nonthermal 
superradiation. However, the Hawking evaporation disappears and only 
the superradiation exists in extremal black holes because the Hawking
 temperature vanishes for extremal black holes.

To further explore the properties of extremal black holes, it is of 
interest to compute other critical exponents. The two-point correlation
 function is a powerful tool to study the critical behavior. Near the 
 critical points, the correlation function has usually the form at a 
 large distance [28-30],
\begin{equation}
G(r)\sim \frac{\exp (-r/\xi)}{r^{d-2+\eta}},
\end{equation}
where $\eta$ is the Fisher's exponent, $d$ is the effective spatial 
dimension of the system under consideration, and $\xi$ is the correlation
 length which is also divergent at critical points. The critical exponents 
 related to the correlation length are defined as
\begin{eqnarray}
\xi &\sim & \varepsilon ^{-\nu}_M \ \ \ ({\rm for\ J\ fixed}),\nonumber\\
    &\sim &\varepsilon ^{-\mu}_J \ \ \ ({\rm for\ M\ fixed}),
\end{eqnarray}
for the BTZ black holes. Combining these critrical exponents $\nu$, $\mu$,
 and $\eta$ with the effective spatial dimension $d$ and  critical
  exponents (42) yields the scaling laws of the ``second kind'',
\begin{equation}
\nu (2-\eta )=\gamma, \ \ \nu d=2-\alpha, \ \ \mu (\beta +\gamma)=\nu.
\end{equation}
Unfortunately, we have not as yet knowledge about the two-point 
correlation function of quantum black holes. To proceed to our 
discussion, here we use the correlation function of a scalar field
 in the black hole background to mimic the one of black holes. The 
 work of Traschen [21] tells us that for the extremal RN black holes, 
 the scalar field has the scaling symmetry and long range correlations, 
 i.e., the effect of the source falls off $y^{-1}$; while for the 
 nonextremal black holes, there is no scaling symmetry, and the 
 influence of  the source falls off exponentially fast, like
  $e^{2\kappa y}$, where $\kappa$ is the surface gravity of black 
  holes and $y$ is the usual tortoise coordinate. Therefore, the
   inverse surface gravity of black holes plays the role of correlation
    length. In the BTZ black holes, this statement is also valid [35,36].
     With the help of (48) and (11), we get
\begin{equation}
\nu=\mu=1/2.
\end{equation}
Substituting (50) into (49), we have
\begin{equation}
\eta=-1, \ \ \ d=1.
\end{equation}
Due to the universality of scaling laws, we surprisedly find that 
the effective dimension of extremal black holes is one, at least 
for the BTZ black holes and Kerr-Newman black holes because their 
corresponding critical exponents are same [17]. The result is in 
agreement with the point of view  that the extremal  black holes 
are the Bogomol'nyi saturated string states,  or more generally, 
the extremal black hole is a special kind of strings [26]. Of course, 
in order to confirm completely the assumption, further investigation 
for extremal black holes is necessary. As to the BTZ black hole, an 
important fact is that the BTZ black hole is also an exact solution 
of string theory in three dimensions [37]. Horowitz and Strominger 
[38] have shown that some extremal black extended solutions in string
 theories are precisly the supersymmetric solutions describing the 
 macroscopic fundamental strings or $p$-branes. Therefore, our 
 conclusion is consistent with these evidences. In the meantime, 
 it seems to imply the critical behavior of extremal black holes 
 could be confirmed in the string theories. In addition to the 
 BTZ black hole solution, the three dimensional string theory has
  the black string solution, dual solution of the BTZ black hole. The 
  black string has the similar
quantum properties as those of the BTZ black hole [39]. So we argue 
that the extremal black string is also a critical point and the
 similar critical behavior will appear in the extremal black strings.

 In addition, in the case of extremal black holes, the correlation 
 function (47) becomes
\begin{equation}
G(r)\sim r^2 \exp (-r/\xi),
\end{equation}
which shows that the radiation of extremal black holes is nonthermal.
 It is again in agreement with the fact that the Hawking evaporation 
 is killed in the extremal black holes and only the nonthermal 
 superradiation is left.

All these discussions manifest that the extremal black hole is a 
critical point, and a phase transition takes place from the extremal 
to nonextremal black holes. In this phase transition point, the scaling
 laws of the ``first kind'' is satisfied automatically, and the 
 ``second kind '' suggests that the effective spatial dimension of
  extremal black holes is one. This is the common feature of extremal
   black  holes. In the BTZ black holes, we find that the massless BTZ
    black hole is also a critical point. This is  a special example in
     the known black  holes. In the  next section, we will discuss the
      case.

\section{The critical point in spinless BTZ black holes}
 
In the spinning BTZ black hole solution (1), if $J=0$, Eq. (1) then 
describes a spinless BTZ black hole, 
\begin{equation}
ds^2=-(-M+\frac{r^2}{l^2})dt^2 +(-M +\frac{r^2}{l^2})^{-1}dr^2 +r^2d\phi ^2.
\end{equation}
The spinless BTZ black holes have only one physical parameter, 
mass $M$, and a horizon $r_+=\sqrt{M}l$. So it has no extremal limit 
in the usual sense. The Hawking temperature of the hole is
 $\beta ^{-1}=\sqrt{M}/(2\pi l)$, the heat capacity is
  $C=4\pi l\sqrt{M}$, and the entropy $S=4\pi l\sqrt{M}$.  
  When $M\rightarrow 0$, all these quantities approach  zero. In 
  the microcanonical ensemble, from Eqs. (23) and (26) we see that 
  the eigenvalue of temperature fluctuation mode is zero and
   $<\delta \beta\delta\beta>$ diverges as $M$ approaches zero. 
   Therefore, the massless BTZ black hole may be also a critical 
   point. In fact, it indeed corresponds to the extremal black holes
    in the usual sense. First, the massless BTZ black hole has zero
     Hawking temperature, zero entropy, and vanishing heat capacity.
      These are the same as the corresponding quantities of usual 
      extremal black holes. Although the usual extremal black holes
       have the nonzero area of horizon, their entropies vanish 
       identically [18,19]. So no difference of thermodynamics 
       exists between the massless BTZ black holes and usual 
       extremal black holes. The only difference is that $M=0$ 
       for the spinless BTZ black hole  in the extremal limit.
        Second, Coussaert and Henneaux [27] have shown that the
         massless BTZ black hole has two exact supersymmetries,
          and the supersymmetry is absent for a generic spinless
           BTZ black hole. This manifests that the massless hole 
           and generic hole are two very different phases. The 
           phase transition takes place from a phase to other 
           phase. Again, the massless hole is in the disordered 
           phase and the generic  spinless BTZ black hole in the
            ordered phase. 
Moreover, the supersymmetry shows the massless BTZ black hole corresponds
 to the usual extremal black hole. Third,  the geometric structure of the
  massless BTZ black hole looks like  the one of extremal RN black holes.
   Both of them have an infinite long throat near their horizons and 
   scaling symmetry. Although the massless BTZ black hole has zero-length
    horizon, the singularity which coincides with the horizon, is invisible
     to an observer at infinity [27]. So, in this paper the massless BTZ 
     solution is referred to as the massless BTZ black hole. Finally, 
     unlike the flat Minkowski spacetime, the background of massless 
     BTZ black holes is not flat, and the Casimir energy exists, which
      results in the quantum instability of the massless BTZ black hole
       [35]. When the reaction effect of quantum fields is taken into 
       account, a non-zero horizon will be developed in the massless 
       BTZ black hole. According to this point, Lifschytz and Ortiz 
       [35] argued that the massless hole is not the end of Hawking 
       evaporation and the end might be the 2+1 dimensional anti-de 
       Sitter spacetime (5). Thus, the massless hole separates the 
        generic spinless BTZ black hole from the anti-de Sitter space.

Combining the above points, we argue that the massless BTZ black hole 
is equivalent to the usual extremal black hole in certain senses, and 
is a critical point of spinless BTZ black holes. A second-order phase 
transition will take place from $M=0$ to $M\ne 0$ black holes because 
the entropy is continuous in this process. 

\section{ Conclusion and discussion}

 In this work we have investigated the  critical behavior and phase 
 transition in the 2+1 dimensional BTZ black holes. By calculating 
 the equilibrium fluctuations of BTZ black holes, we have found that
  some second moments are divergent for extremal BTZ black holes in 
  the microcanonical ensemble and canonical ensemble, but all second
   moments are finite in the grand canonical ensemble even under the
    extremal condition. The divergence of second moments in the
     microcanonical ensemble shows the extremal BTZ black hole 
     is the critical point of the equilibrium sequence of BTZ
      black holes. Some critical exponents have been calculated and 
      they satisfy the scaling laws of the ``first kind''. By using 
      the correlation function of scalar fields in the black hole 
      background to replace the one of quantum black  holes, the 
      effective spatial dimension of extremal black holes is found 
      to be one. This is surprisingly in agreement with the assumption 
      that the extremal black holes are the Bogomol'nyi saturated strings
       states  and the extremal black hole can be regarded as the 
       elementary particles like states in string theories, 
       although we do not as yet know whether there exist the
        essential relations between our result and the equivalent 
        assumption of extremal black holes and  saturated string 
        states. In addition, we have argued that the massless BTZ 
        black hole is a critical point of the spinless BTZ black 
        hole, and a second-order phase transition will take place
         from the extremal to nonextremal BTZ black holes. Because 
         of the similarity of thermodynamics between BTZ black holes
          and three dimensional black strings [39], we believe that 
          the similar critical behavior also appears in the black 
          strings.

 To better understand the critical behavior of extremal black holes,
  further investigation is needed. For example, are the critrical
   exponents universal? Is the effective spatial dimension of other
    extremal black holes also one? Could we obtain a reduced quantum 
    model of black holes, in which the critical behavior can be 
    confirmed? These issues are currently under investigation.

Finally, we make a comment on the phase transition of black holes.
 First, according to the second and third laws of black hole 
 thermodynamics, an extremal black hole cannot be reached by 
 a physical process. The extremal black hole, however, can be
  produced by the pair creation [18]. Therefore, the second 
  and third laws cannot exclude the phase transition from the
   extremal black holes to nonextremal black holes. Second, 
   if the entropy of extremal black holes still satisfies the 
   Bekenstein-Hawking area formula, the phase transition is
    second-order because the entropy is continuous from the 
    extremal to nontremal black holes. Instead if one accepts 
    the view of point that the entropy of extremal black holes
     vanishes identically, the phase transition is higher than 
     second-order, because in this case the entropy is d
     iscontinuous. But, the phase transition in the spinless 
     BTZ black holes is still a second-order one because of 
     the continuity of entropy.

\begin{flushleft}
{\bf \large Acknowledgments}
\end{flushleft}

The work of R.G.C. was supported in part by China Postdoctoral 
Science Foundation, he would like to thank O. Kaburaki for sending 
some copies of his recent works to him, and Dr.
 Y. K. Lau for useful discussions.

\end{document}